\documentclass[11pt]{article}
\usepackage[T1]{fontenc}
\usepackage[utf8]{inputenc}
\usepackage[english]{babel}
\usepackage{lmodern}
\usepackage{microtype}
\usepackage{mathtools}
\usepackage{amssymb}
\usepackage{braket}
\usepackage[dvipsnames]{xcolor}
\definecolor{darkblue}{rgb}{0,0,0.6}
\definecolor{purple}{rgb}{0.4,0.2,0.7}
\usepackage[margin=2.7cm]{geometry}
\usepackage[toc,page]{appendix}
\usepackage[
  hyperfootnotes=false,
  colorlinks=true,
  linkcolor=darkblue,
  citecolor=purple,
  urlcolor=darkblue
]{hyperref}

\hypersetup{
  pdftitle={De Sitter Unitarity versus Reflection Positivity},
  pdfauthor={Yu-ki Suzuki},
  pdfsubject={De Sitter representations, reflection positivity, and holography},
  pdfkeywords={de Sitter space, reflection positivity, holography, complementary series, Osterwalder-Schrader reconstruction}
}

\allowdisplaybreaks
\numberwithin{equation}{section}

\newcommand{\be}{\begin{equation}}
\newcommand{\ee}{\end{equation}}
\newcommand{\nn}{\nonumber}
\newcommand{\dS}{\mathrm{dS}}
\newcommand{\AdS}{\mathrm{AdS}}
\newcommand{\cH}{\mathcal H}
\newcommand{\cE}{\mathcal E}
\newcommand{\cN}{\mathcal N}
\newcommand{\R}{\mathbb R}
\newcommand{\C}{\mathbb C}
\newcommand{\OS}{\mathrm{OS}}
\newcommand{\KG}{\mathrm{KG}}
\newcommand{\dd}{\mathrm d}
\newcommand{\ip}[2]{\left\langle #1,#2\right\rangle}
\newcommand{\norm}[1]{\left\|#1\right\|}
\DeclareMathOperator{\Span}{span}

\begin{document}

\thispagestyle{empty}
  \begin{flushright}
  RIKEN-iTHEMS-Report-26
    \end{flushright}
    
    \bigskip
\begin{center}
\vspace*{6mm}

{\huge\bfseries Hilbert space relation from dS to AdS through Reflection Positivity}

\vspace{0.6in}

{\Large\bfseries Yu-ki Suzuki}

\vspace{0.5in}

{\large
RIKEN Center for Interdisciplinary Theoretical\\
and Mathematical Sciences (iTHEMS),\\
Wako, Saitama 351-0198, Japan
}

\vspace{0.5in}
\end{center}

\begin{abstract}
De Sitter unitarity and  reflection positivity are logically independent. We study their relation for a free scalar field on a fixed de Sitter background,  restricting attention to the one-particle sector. Motivated by Witten's de Sitter pairing and the radial adjoint of Bousso, Maloney and Strominger, we compose the invariant Hilbert product with equatorial reflection. In the standard principal-series realization, 
this form vanishes identically and its Osterwalder-Schrader quotient contains no states. For the complementary series, the reflected  product becomes a nonzero reflected kernel on the unit ball. A  result of Neeb and Ólafsson shows that this kernel is positive semidefinite precisely when the lower  weight satisfies the scalar conformal unitarity bound. The resulting Hilbert space with the reflected inner product furnishes a positive-energy representation of AdS isometries according to Neeb and Ólafsson. We identify that the  range coincides with the Klebanov-Witten alternative-quantization window from  physics point of view. This correspondence suggests a novel relation from the representation viewpoint between AdS and dS without the conventional analytic continuation of radius.
\end{abstract}

\vspace{1in}

\pagebreak

\renewcommand{\contentsname}{\texorpdfstring{\textcolor{darkblue}{Contents}}{Contents}}
{\large
\tableofcontents
}
\clearpage
\section{Introduction and summary}
\label{sec:introduction}

The holographic principle is one of the central  ideas in the
search for a quantum  gravity
\cite{tHooft:1993dmi,Susskind:1994vu}.  Its best-understood realization is
the AdS/CFT correspondence \cite{Maldacena:1997re}, which relates quantum
gravity in an asymptotically anti-de Sitter spacetime to a conformal field
theory in one lower dimension.  By contrast, a comparably precise
holographic description of de Sitter space remains elusive.  The pioneering
dS/CFT proposal associates quantum gravity in asymptotically de Sitter space
with a Euclidean conformal field theory at future  infinity
\cite{Strominger:2001pn}.  A useful strategy for exploring this proposal is
to utilize the analytic continuation of the cosmological constant or the radius of  curvature \cite{Maldacena:2002vr}.

In this paper, we investigate a different representation-theoretic relation
between dS and AdS structures.   We consider a free
scalar field on a fixed $\dS_{d+1}$ background and restrict attention to its
one-particle sector, which carries a scalar unitary irreducible
representation of $G=SO_0(1,d+1)$
\cite{Sun:2021thf,Sengor:2019mbz,Anous:2020nxu}.  Basically, we  interpret the reflection-positivity results of Neeb and
\'{O}lafsson \cite{neeb2014reflection,neeb2018reflection} from a physics point of view and explain the relation to the  AdS result.

 The key step in their formulation  is the reflection along the equator of $S^d$ geometrically associated with the spatial sphere in global de Sitter coordinates. A similar reflection was also
motivated by Witten's de Sitter pairing \cite{Witten:2001kn} and the radial
adjoint of Bousso, Maloney, and Strominger \cite{Bousso:2001mw}.
 Utilizing the stereographic projection, the reflection turns it into radial inversion on the unit ball, which is realized in the radial quantization of CFT.
Composing the de Sitter product with this reflection defines a second
sesquilinear form.  Reflection positivity requires this form to be positive
semidefinite.  Because it may nevertheless vanish identically, we separately
require a nonzero Osterwalder-Schrader quotient.  The resulting novel Hilbert space can be understood as the AdS Hilbert space.

We apply this  to  scalar unitary series in dS.  In the standard
principal-series realization, the de Sitter product is local, so a
hemisphere-supported wavefunction pairs trivially with its reflection.  The
form vanishes identically and its OS quotient is zero.  For the
complementary series, the de Sitter product is a nonlocal shadow product.
Writing the real weights as $\Delta_\pm=d/2\pm\nu$, we use states in a position basis of
weight $\Delta_-$ and smearing wavefunctions of the dual weight $\Delta_+$.
Equatorial reflection then produces the  kernel of
dimension $\Delta_-$.

The Neeb--\'{O}lafsson classification implies that, within the open
complementary series, this kernel is positive semidefinite precisely when
$\nu\leq1$, or equivalently when $\Delta_-$ satisfies the scalar conformal
unitarity bound.  The necessity of this condition has a simple physical
meaning. The reflected norm of the level-two trace descendant changes sign
at $\nu=1$.  

The quotient also changes the real form represented unitarily.  Equatorial
reflection splits the de Sitter algebra into reflection even and odd parts.  The
fixed subgroup and the conformal compression semigroup act directly on the
positive region.  Multiplying the reflection-odd generators by $i$
reverses the sign of their mutual commutator and produces the $c$-dual
algebra $\mathfrak{so}(2,d)$. A similar construction in the Klein-Gordon inner product in the static patch was analyzed in \cite{Jafferis:2013qia}. The resulting nontrivial OS
quotient is therefore a scalar positive-energy lowest-weight module with the 
lowest energy $\Delta_-$.

Finally, identifying the de Sitter and AdS mass parameters  turns 
$0<\nu<1$ into the Klebanov--Witten alternative-quantization window
\cite{Klebanov:1999tb}.  This agreement concerns representation parameters
and Hilbert-space structures, not 
an analytic continuation of the radius of  curvature  between AdS and dS.  The
construction produces a single global scalar conformal module from a free
one-particle sector. 

Section~\ref{sec:origin} motivates the reflection along the equator of the sphere , and
Section~\ref{sec:representations} reviews the scalar unitary irreducible representations.
Section~\ref{sec:positivity} performs the positivity test based on the mathematical theorem.
Section~\ref{sec:reconstruction} identifies the $c$-dual algebra and positive
radial Hamiltonian, while Section~\ref{sec:kw} explains the relation to AdS physics or 
Klebanov-Witten window.  Appendix~\ref{app:kg} connects the Klein-Gordon inner product and the unitary irreducible representations.

\section{From de Sitter pairings to radial reflection}
\label{sec:origin}

\subsection{Witten's Hermitian pairing}

A conventional S-matrix is not available in global de Sitter space.  One reason is the absence of a Killing vector field that is timelike everywhere. Hence, we do not have  a preferred decomposition into positive and negative-frequency modes.  Another reason is that cosmological horizons prevent any single observer from accessing the entire spacetime.  Past
and future conformal infinity, denoted $\mathcal I^-$ and $\mathcal I^+$,
are spacelike and  no observer has access to all boundary data on both components.

Nevertheless, the gravitational path
integral may define a transition amplitude between asymptotic boundary data.
Let $\cH_i$ and $\cH_f$ denote the spaces of initial data on $\mathcal I^-$
and final data on $\mathcal I^+$, respectively.  Witten proposed treating the
transition amplitude as a bilinear map \cite{Witten:2001kn}
\be
\mathcal M:\cH_f\otimes\cH_i\longrightarrow\C,
\quad 
\mathcal M(f,i)=\langle f|i\rangle.
\label{eq:Witten-bilinear}
\ee
Here $\ket{i}\in\cH_i$ and $\bra{f}\in\cH_f$, and $\mathcal M(f,i)$ is linear in both
arguments. We can project out the null vectors by taking  quotients. This map does not define a norm purely on $\cH_f$ or $\cH_i$ since
its two arguments belong to different spaces $\cH_f$ and $\cH_i$.  CPT conjugation supplies an antilinear map
$\Theta:\cH_i\to\cH_f$ between initial and final data and hence converts the
bilinear amplitude into a sesquilinear pairing on $\cH_i$,
\be
(j,i)_W=\mathcal M(\Theta j,i)=\langle\Theta j|i\rangle.
\label{eq:Witten-pairing}
\ee
Witten conjectured positivity of this pairing in quantum gravity.  Our analysis is more modest.  We keep only the structural lesson that a de Sitter inner product may involve a reflection.  We then study the corresponding question in a fixed-background scalar one-particle representation.

It is important not to identify \eqref{eq:Witten-pairing}  with the ordinary CFT radial norm.  CPT contains an antipodal operation exchanging $\mathcal I^-$ and $\mathcal I^+$, whereas radial quantization on one copy of $S^d$ uses reflection across an equator.  The relation between these operations depends on the parity convention.

\subsection{The radial adjoint of Bousso, Maloney and Strominger}

 Let us consider global dS$_3$
 \be
 ds^2=-d\tau_{\dS}^2+\cosh^2\tau_{\dS}(d\vartheta^2+\sin^2\vartheta d\varphi^2),
 \ee
 and focus on the scalar field.
Bousso, Maloney and Strominger chose the time-reversal and parity operations
\cite{Bousso:2001mw}
\be
\mathsf T:\tau_{\dS}\mapsto-\tau_{\dS},
\quad 
\mathsf P:(\vartheta,\varphi)\mapsto(\vartheta,\varphi+\pi).
\label{eq:BMS-PT}
\ee
If $\Omega=(\vartheta,\varphi)$ and
$\Omega_A=(\pi-\vartheta,\varphi+\pi)$ is its antipodal point on $S^2$, then
\be
\mathsf P\Omega_A=(\pi-\vartheta,\varphi).
\ee
The combined operation is therefore reflection across the equator, not the
antipodal map on the same sphere.  In their free-field realization,
$\phi^{\rm in/out}_\pm$ denotes the boundary coefficient of one of the two
scalar falloffs at past/future infinity.  The boundary adjoint takes the form
\be
\left(\phi^{\rm in}_\pm(\Omega)\right)^\dagger
=
\phi^{\rm out}_\pm(\mathsf P\Omega_A).
\label{eq:BMS-adjoint}
\ee

We generalize the above argument to $S^d$. 
We label the coordinates as
\be
\Omega=(\Omega_\perp,\boldsymbol\Omega)\in S^d,
\quad 
\Omega_\perp^2+|\boldsymbol\Omega|^2=1,
\ee
and define the equatorial reflection
\be
r_0(\Omega_\perp,\boldsymbol\Omega)
=(-\Omega_\perp,\boldsymbol\Omega).
\label{eq:equatorial-reflection}
\ee
Using stereographic projection $s:\R^d\to S^d$ from the south pole,
\be
s(x)=
\left(\frac{1-|x|^2}{1+|x|^2},
\frac{2x}{1+|x|^2}
\right),
\quad  x\in\R^d,
\label{eq:stereographic}
\ee
one finds
\be
s^{-1}r_0s(x)=\frac{x}{|x|^2}.
\label{eq:radial-inversion}
\ee
The subscript $\perp$ denotes the coordinate normal to the chosen equator.
The positive hemisphere $\Omega_\perp>0$ is mapped to the unit ball
$B_1=\{x\in\R^d:|x|<1\}$.  Thus, the equatorial reflection exchanges the inside
and outside of the ball and becomes the inversion used to define the adjoint
in radial quantization.  This observation explains why a de Sitter reflection
can test a CFT-like radial norm.  

\section{Scalar unitary representations of \texorpdfstring{$SO_0(1,d+1)$}{SO(1,d+1)}}
\label{sec:representations}

\subsection{Algebra and de Sitter reality}

We set $G=SO_0(1,d+1)$, the identity component of the de Sitter isometry
group, and write $\mathfrak g=\mathfrak{so}(1,d+1)$ for its Lie algebra.
The (d+1)-dimensional de Sitter space is realized by the hypersurface in the (d+2)-dimensional Minkowski space 
\begin{align}
    -X_0^2+X_1^2+\cdots+X_{d+1}^2=1,\nn\\
    ds^2=-dX_0^2+dX_1^2+\cdots+dX_{d+1}^2,
\end{align}
where we set the dS radius to $1$.
We realize
$\mathfrak g$ by  generators
\be
L_{AB}=X_A\partial_B-X_B\partial_A,\quad L_{AB}=-L_{BA},\quad A,B=0,1,\cdots,d+1,
\ee
 satisfying
\be
[L_{AB},L_{CD}]
=\eta_{BC}L_{AD}-\eta_{AC}L_{BD}
+\eta_{AD}L_{BC}-\eta_{BD}L_{AC},
\quad 
\eta={\rm diag}(-1,+1,\ldots,+1).
\label{eq:dS-algebra}
\ee
We can reorganize the  generators in
the Euclidean conformal basis
\begin{align}
L_{ij}&=M_{ij},
&L_{0,d+1}&=D,
\nn\\
L_{d+1,i}&=\frac12(P_i+K_i),
&L_{0,i}&=\frac12(P_i-K_i),
\label{eq:conformal-basis}
\end{align}
where $i,j=1,\cdots,d$.
We use anti-Hermitian Lie-algebra generators on the de Sitter Hilbert space,
\be
L_{AB}^{\dagger_{\dS}}=-L_{AB}.
\label{eq:dS-reality}
\ee
This is different from the usual CFT adjoint. Usual CFT conjugation satisfies $P^\dagger=K$, but here we have $P^\dagger=-P$. 

Let us review the representation of UIR of $SO_0(1,d+1)$. 
We write $\ket{\Delta,x}$ for the distribution vector labeled by
$x\in\R^d$ and conformal weight $\Delta$. Notice that we consider the Hilbert space later, but the basis itself is divergent. This is the same situation in quantum mechanics, where the $\ket{x}$ is not normalizable and should be regarded as a distribution. The vector at the origin obeys the
stabilizer (primary-like) conditions
\be
K_i\ket{\Delta,0}=0,
\quad
D\ket{\Delta,0}=\Delta\ket{\Delta,0},\quad M_{ij}\ket{\Delta,0}=0,
\label{eq:primary-like}
\ee
and vectors at other points are generated by translations,
\be
\ket{\Delta,x}=e^{x\cdot P}\ket{\Delta,0}.
\label{eq:translated-position-state}
\ee
The wavefunction $f(x)$ produces a normalizable  smooth smearing
\be
\ket{f}=\int_{\R^d}\dd^dx\,f(x)\ket{\Delta,x}.
\label{eq:smeared-state}
\ee
Below,  we assume that the distribution vector
has the lower weight $\Delta=\Delta_-$, whereas the smearing profile
transforms with the dual weight $d-\Delta_-=\Delta_+$.    Notice that the
real generalized eigenvalue in \eqref{eq:primary-like} does not contradict the
skew-adjointness of $D$ on the de Sitter Hilbert space since
$\ket{\Delta,0}$ is not a normalizable  eigenvector in the Hilbert space.

\subsection{Mass parameters and boundary weights}

For a scalar field $\Phi$ of mass $m_{\dS}$ on $\dS_{d+1}$,
\be
(\Box_{\dS}-m_{\dS}^2)\Phi=0.
\ee
Introduce the spectral parameter
\be
\nu=\sqrt{\frac{d^2}{4}-m_{\dS}^2}.
\label{eq:nu-dS}
\ee
For $m_{\dS}^2<d^2/4$, the parameter $\nu$ is real.  The (open) complementary series is in the range
\be
0<\nu<\frac d2,
\quad 
\Delta_\pm=\frac d2\pm\nu,
\label{eq:comp-weights}
\ee
In planar coordinates, a
late-time solution has
the two leading falloffs
\be
\Phi(\eta,x)
\sim
(-\eta)^{\Delta_-}\phi_-(x)
+(-\eta)^{\Delta_+}\phi_+(x),
\quad
\eta\to0^-.
\label{eq:dS-falloffs-comp}
\ee
The coefficient  $\phi_-$ and $\phi_+$ transform with the two
shadow-related weights $\Delta_-$ and $\Delta_+$, respectively. 

 For
$m_{\dS}^2>d^2/4$, it is convenient to write $\nu=i\mu$ with
$\mu>0$ and  this is the scalar principal series,
\be
\nu=i\mu,
\quad 
\mu>0,
\quad 
\Delta_\pm=\frac d2\pm i\mu.
\label{eq:principal-weights}
\ee
In general, the label $\Delta$ appeared in Unitary irreducible representation (UIR) and the conformal weight of CFT in the dS/CFT should be distinguished \cite{Hogervorst:2021uvp}. Only in the free field case, these labels coincidentally match.

\subsection{Inner products and Hilbert space}

  For the principal series, the inner product is given by the $L^2$ norm
\be
\cE_\mu=L^2(S^d),
\quad
\ip{f}{g}_{\rm \mu}
=
\int_{S^d}\dd\Omega\,f(\Omega)^*g(\Omega).
\label{eq:principal-product}
\ee
The $\mu$ labels the principal series and $\Omega$ is the coordinate on $S^d$. This sphere can be regarded as the one on a constant global timeslice
\be
ds^2=-dt^2+\cosh^2t d\Omega_d^2.
\ee
In the principal series, the weight of $f(\Omega)^*f(\Omega)$  is $\Delta+\Delta^*=d$, which makes this local inner product conformally invariant, since it cancels the factor from the measure.

For the complementary series, the local $L^2$ pairing is not invariant under the de Sitter isometry
because the weight is real rather than satisfying $\Delta+\Delta^*=d$.
Invariance is restored by inserting the shadow kernel.  Up to a positive
normalization, one can define
\be
\ip{f}{g}_{\nu}
=
\int_{S^d}\dd\Omega
\int_{S^d}\dd\Omega'\,
 f(\Omega)^*Q_\nu(\Omega,\Omega')g(\Omega'),
\label{eq:comp-product}
\ee
with
\be
Q_\nu(\Omega,\Omega')
=
(1-\Omega\cdot\Omega')^{\nu-d/2}
=
(1-\Omega\cdot\Omega')^{-\Delta_-},
\label{eq:shadow-kernel-sphere}
\ee
where $\nu$ labels the complementary series.  
In flat coordinates, this becomes the shadow inner product
\be
\ip{f}{g}_{\nu}
\propto
\int_{\R^d}\dd^dx
\int_{\R^d}\dd^dy\,
 f(x)^*\frac{1}{|x-y|^{2\Delta_-}}g(y),
\label{eq:shadow-product-flat}
\ee
with the conformal factors from stereographic projection absorbed into the
wavefunctions.  These functions transform with weight
$\Delta_+=d-\Delta_-$.  

We also have other branches, such as discrete series and exceptional series, in the unitary irreducible representation.
In our paper, we focus on the principal and complementary series below.

Both \eqref{eq:principal-product} and \eqref{eq:comp-product} arise from the bulk Klein-Gordon form after restricting to a positive-frequency one-particle subspace.  A short derivation is collected in Appendix~\ref{app:kg}. See \cite{Guijosa:2003ze,Anous:2020nxu}, for earlier works.  

\section{Equatorial reflection and positivity}
\label{sec:positivity}
In this section, we compose the de Sitter inner product with equatorial reflection and restrict the resulting form to profiles supported in the positive hemisphere. We then determine the mass range in which this reflected form is positive semidefinite.

\subsection{Reflected form}
First let us define the reflection along the equator.
Let
\be
S^d_+=\{\Omega\in S^d:\Omega_\perp>0\}
\ee
be the open positive hemisphere.  Let $\theta$ denote the involution on
the wavefunction induced by \eqref{eq:equatorial-reflection},
\be
(\theta f)(\Omega)=f(r_0\Omega).
\label{eq:theta-action}
\ee
We define the space of smooth wavefunctions, which has a support on the positive hemisphere $S^d_+$ as  $\mathcal{D} _+$.   We take  the closure of the Hilbert space of UIR corresponding to the positive-region
Hilbert subspace by
\be
\cE_{\kappa,+}
:=
\overline{\mathcal{D}_+}^{\,\norm{\cdot}_\kappa}
\subset \cE_\kappa ,
\ee
where the closure is taken in the original de Sitter Hilbert norm
$\norm{\cdot}_\kappa$ and $\kappa$ denotes the relevant representation parameter, namely $\mu$ in the principal series and $\nu$ in the complementary series.
  We define the reflected form by
\be
\ip{u}{v}_{\theta,\kappa}
:=
\ip{u}{\theta v}_{\kappa},
\quad 
u,v\in\cE_{\kappa,+}.
\label{eq:theta-product}
\ee
This is a new form on the positive-region subspace, not the original de
Sitter inner product.  It is Hermitian because $\theta$ is a unitary
involution. 
We would like to derive the parameter region, where this inner product becomes positive semidefinite.

\subsubsection*{The principal series branch}

We take $\kappa=\mu$. The inner product in the principal series in UIR of de Sitter space is given by the $L^2$ norm. Then, for $u,v\in\cE_{\mu,+}$, the support of $\theta v$ lies in the negative
hemisphere.  It is disjoint from the support of $u$ up to the equator, which
has measure zero. Thus, the integral should vanish in the  principal series  
\be
\ip{u}{v}_{\theta,\mu}
=
\ip{u}{\theta v}_\mu
=0
\quad 
\text{for all }u,v\in\cE_{\mu,+}.
\label{eq:principal-zero}
\ee

\subsubsection*{The complementary series branch}

We next evaluate the the complementary series branch. We take
$\kappa=\nu$ and  the original inner product is given by the shadow product
\eqref{eq:comp-product}. 

Since $u$ and $v$ are supported in $S^d_+$, inserting the definition of the
shadow inner product and changing the reflected integration variable gives
\be
\ip{u}{v}_{\theta}
=
\int_{S^d_+}\dd\Omega
\int_{S^d_+}\dd\Omega'\,
 u(\Omega)^*Q_\nu(r_0\Omega,\Omega')v(\Omega').
\label{eq:theta-product-sphere}
\ee
To compare this expression with radial quantization, we move to flat coordinates by stereographic map.  A direct calculation using
\eqref{eq:stereographic} gives
\be
1-r_0s(x)\cdot s(y)
=
\frac{2\left(1-2x\cdot y+|x|^2|y|^2\right)}{(1+|x|^2)(1+|y|^2)}.
\label{eq:reflected-distance}
\ee
The denominator in \eqref{eq:reflected-distance} is positive.    Explicitly, the
stereographic measure is transformed into
\be
\dd\Omega
=
\left(\frac{2}{1+|x|^2}\right)^d\dd^dx.
\ee
Since $d-\Delta_-=\Delta_+$, the wavefunction also transforms as
\be
\widetilde u(x)
:=
(1+|x|^2)^{-\Delta_+}u(s(x)),
\label{eq:ball-profile-rescaling}
\ee
and similarly for $v$.  Up to the irrelevant overall positive constant
$2^{2d-\Delta_-}$, all one-point factors are now contained in
$\widetilde u$ and $\widetilde v$.  Dropping the tildes from this point, the reflected form on $B_1$ (ball) becomes
\be
\ip{u}{v}_{\theta}
=
\int_{B_1}\dd^dx
\int_{B_1}\dd^dy\,
 u(x)^*R_\nu(x,y)v(y),
\label{eq:theta-product-ball}
\ee
where
\be
R_\nu(x,y)
=
\left(1-2x\cdot y+|x|^2|y|^2\right)^{-\Delta_-}.
\label{eq:R-kernel}
\ee
 
To see the CFT interpretation, consider a scalar primary
$\mathcal O$ of dimension $\Delta$ with two-point function
$\langle\mathcal O(x)\mathcal O(y)\rangle=|x-y|^{-2\Delta}$.  Radial conjugation
acts as
\be
\mathcal O(x)^\dagger
=
|x|^{-2\Delta}\mathcal O\left(\frac{x}{|x|^2}\right).
\ee
Therefore the overlap of the states created at $x,y\in B_1$ is
\be
\left\langle\mathcal O(x)^\dagger\mathcal O(y)\right\rangle
=
\left(1-2x\cdot y+|x|^2|y|^2\right)^{-\Delta}.
\ee
Taking $\Delta=\Delta_-$ gives exactly $R_\nu(x,y)$. Thus, our reflection is the same as the one used in the radial quantization.

\subsection{Positivity of the reflected form}

In this section, we explain when the reflected form becomes positive semidefinite in the complementary series. We first introduce a mathematical theorem and later see physical derivation of this result. 
For the rigorous derivation, please refer to 
\cite{neeb2014reflection,neeb2018reflection}.

\subsubsection*{Positive semidefinite range}
We will introduce the theorem by Neeb-\'{O}lafsson \cite{neeb2014reflection,neeb2018reflection}. 

For $d\geq2$ and $0<\nu<d/2$,  the kernel \eqref{eq:R-kernel} is positive semidefinite on the open unit
ball if and only if
\be
\nu\leq1.
\label{eq:nu-window}
\ee
 Correspondingly, the reflected form is positive definite if and only if
\be
\frac{d^2}{4}-1
\leq
m_{\dS}^2
<
\frac{d^2}{4}.
\label{eq:dS-mass-window}
\ee
This is a proper subrange of the complementary series.  For $d=2$, every representation in the open complementary series,
\be
0<m_{\dS}^2<1,
\ee
is reflection positive. 

\subsubsection*{Physical interpretation}

Before delving into the intuitive interpretation, we would like to notice that the restriction above is  not a new bulk unitarity bound.  It answers a different question: whether the same one-particle data can be interpreted as a positive radial-quantization sector after equatorial reflection. 

From physics point of view, we cannot derive necessary and sufficient conditions, but we will derive at least necessary condition. The derivation is similar to the derivation of unitarity bound in unitary CFTs.

Let us denote $\Delta=\Delta_-$ and
\be
R_\Delta(x,y)
=
\left(1-2x\cdot y+|x|^2|y|^2\right)^{-\Delta}.
\ee
The   trace of the descendant-like states at level two is proportional to
\be
\left.
\partial_x^2\partial_y^2R_\Delta(x,y)
\right|_{x=y=0}
=
8d\Delta\left(\Delta-\frac{d-2}{2}\right).
\label{eq:level-two-norm}
\ee
For $\Delta=d/2-\nu$, this becomes
\be
8d\left(\frac d2-\nu\right)(1-\nu).
\label{eq:level-two-nu}
\ee
Thus, we obtain the desired condition.

\subsection{The quotient Hilbert space}

The reflected form  \eqref{eq:theta-product-sphere} in the complementary series is positive semidefinite.  It may
nevertheless contain nonzero null vectors.  We define its radical or null space by  
\be
\cN_\nu
=
\left\{
 u\in\cE_{\nu,+}:
 \ip{u}{v}_{\theta,\nu}=0
 \text{ for every }v\in\cE_{\nu,+}
\right\}.
\label{eq:null-space}
\ee
 Quotienting by $\mathcal N_\nu$ removes these null directions.  If
$[u]=u+\cN_\nu$ denotes the equivalence class of $u$, the reconstructed
Hilbert space is
\be
\widehat{\cE}_\nu
=
\overline{\cE_{\nu,+}/\cN_\nu}^{\,\norm{\cdot}_{\theta,\nu}},
\quad
\ip{[u]}{[v]}_{\OS,\nu}:=\ip{u}{v}_{\theta,\nu},
\quad
\norm{[u]}_{\theta,\nu}^2=\ip{u}{u}_{\theta,\nu},
\label{eq:OS-Hilbert}
\ee
where we take the Hilbert space completion. We call the resulting positive-definite inner product the Osterwalder–Schrader (OS) inner product.
The OS-form is independent of the chosen
representatives because changing either representative by an element of
$\cN_\nu$ does not change the reflected form.
The bar denotes completion because a Cauchy sequence of equivalence classes
need not converge to a smooth hemisphere-supported wavefunction.

If we consider the set of Null vectors in the principal series, it is given by whole space $\cE_{\mu,+}$ and the resulting positive definite space becomes empty.

\section{ Relation to AdS space}
\label{sec:reconstruction}

The reflection also changes the adjoint.  We write $A^{\dagger_\theta}$ for the
adjoint with respect to the reflected form.  Let $A$ be an operator acting on $u,v\in\mathcal{E}_{\nu,+}$.
The adjoint rule in the reflected form is  explicitly determined
\begin{align}
\ip{Au}{v}_\theta
&=\ip{Au}{\theta v}_\nu
=\ip{u}{A^{\dagger_{\dS}}\theta v}_\nu
\nn\\
&=\ip{u}{\theta(\theta A^{\dagger_{\dS}}\theta)v}_\nu
=\ip{u}{(\theta A^{\dagger_{\dS}}\theta)v}_\theta.
\end{align}
Therefore,
\be
A^{\dagger_\theta}=\theta A^{\dagger_{\dS}}\theta.
\label{eq:theta-adjoint}
\ee
This identity is the algebraic origin of the change from de Sitter to AdS
reality conditions.
 We will examine this point in more detail.

\subsection{The \texorpdfstring{$c$-dual}{c-dual} algebra}

We now determine which real Lie algebra is unitary after the
adjoint has been twisted by reflection.  In the (d+2)-dimensional
embedding space, choose the coordinate $X^{d+1}$ to be the spatial direction
reversed by the equatorial reflection. 

Let $R_0$ denote the ambient lift of
the  reflection $r_0$, so that $R_0$ acts as
$X^{d+1}\mapsto-X^{d+1}$.   The 
conjugation by the reflection defines an  automorphism
\be
\tau(g)=R_0gR_0^{-1},
\quad 
\tau^2=1.
\label{eq:group-involution}
\ee
We use the same symbol $\tau$ for the differential of this automorphism at
the identity.  Every generator then splits uniquely into a reflection-even
and a reflection-odd part,
\be
\mathfrak g=\mathfrak h\oplus\mathfrak q,
\quad 
\mathfrak h=\{X:\tau X=X\},
\quad 
\mathfrak q=\{Y:\tau Y=-Y\},
\label{eq:hq-decomp}
\ee
where $\mathfrak h$ and $\mathfrak q$ are the $+1$ and $-1$ eigenspaces of
$\tau$, respectively.  Since $\tau$ preserves Lie brackets, the parity of a
commutator is the product of the parities of its arguments:
\be
[\mathfrak h,\mathfrak h]\subset\mathfrak h,
\quad 
[\mathfrak h,\mathfrak q]\subset\mathfrak q,
\quad 
[\mathfrak q,\mathfrak q]\subset\mathfrak h.
\ee
Notice that although $\mathfrak h$ forms the Lie algebra, $\mathfrak q$ does not form it. The element in $\mathfrak q$ maps the domain of $u\in \mathcal{E}_\nu^+$ into south hemisphere and thus it is nontrivial that it is well defined after taking quotient by the null space, where the reflected form degenerates. To prevent it, we define the compression semigroup $\Gamma_+$ below.

Let $U_\nu$ denote the original unitary representation of $G$ on
$\cE_\nu$. Below, we write it simply as $U$ when the parameter is fixed.
 The subgroup,
\be
H=(G^\tau)_0,
\quad 
G^\tau=\{g\in G:\tau(g)=g\},
\ee
preserves the equator. Thus, the action of $\mathfrak h$ is well defined on the positive hemisphere.  To discuss the action of $\mathfrak q$, we need to define a compression. In the ball picture, the conformal compression
semigroup $\Gamma_+$ consists of conformal transformations that map $B_1$
into itself
\be
\Gamma_+
:=
\{g\in G\mid g(B_1)\subseteq B_1\},
\quad 
B_1=\{x\in\mathbb R^d:|x|<1\}.
\label{eq:compression-semigroup}
\ee
 The inverse of an element
need not preserve $B_1$ and  it is therefore a semigroup rather than a
group.

Both $H$ and $\Gamma_+$ act directly on the positive region.  The semigroup carries the anti-involution
\be
g^\sharp:=\tau(g)^{-1},
\quad  g\in\Gamma_+.
\label{eq:sharp-involution}
\ee
The compression semigroup is stable under $\sharp$.  For
$g\in\Gamma_+$ and $u,v\in\cE_{\nu,+}$, unitarity of the original
representation and the reflection relation give
\be
\ip{U(g)u}{v}_\theta
=
\ip{u}{U(g^\sharp)v}_\theta.
\label{eq:prequotient-semigroup-adjoint}
\ee
To verify this identity, move $U(g)$ through the original de Sitter inner
product as $U(g^{-1})$.  The reflection relation then gives
$U(g^{-1})\theta=\theta U(\tau(g)^{-1})=\theta U(g^\sharp)$, which restores
the reflected form and produces precisely the right-hand side of
\eqref{eq:prequotient-semigroup-adjoint}.
Because $g^\sharp$ also preserves the positive-region space, this identity
implies $U(g)\cN_\nu\subset\cN_\nu$.  The quotient action is therefore
well defined by
\be
\widehat U(g)[u]:=[U(g)u],
\quad  g\in\Gamma_+,
\label{eq:quotient-action}
\ee
and satisfies the OS adjoint relation
\be
\ip{\widehat U(g)[u]}{[v]}_{\rm OS}
=
\ip{[u]}{\widehat U(g^\sharp)[v]}_{\rm OS},\quad 
\widehat U(g)^{\dagger_{\rm OS}}=\widehat U(g^\sharp).
\label{eq:semigroup-adjoint}
\ee
This OS adjoint $^\dagger_{\rm OS}$ is different from the one with reflection $^\dagger_{\theta}$ since we also take into account the quotient by the null vector space.

\paragraph{Reflected adjoints and continuation to the $c$-dual.}
For $X\in\mathfrak h$, the one-parameter group $\exp(tX)$ lies in the fixed
subgroup $H$ and acts unitarily on the OS quotient.  Differentiating this
quotient action on $\mathcal D_{\OS}$ gives
\be
\ip{Xu}{v}_{\OS}=-\ip{u}{Xv}_{\OS},
\quad 
X\in\mathfrak h,
\quad 
u,v\in\widehat{\mathcal {E}}^d_\nu=\{h\in \widehat{\mathcal {E}}:\lim_{t\to0}:\frac{U(e^{tX})h-h}{t}\,\,\text{exists}\}
\label{eq:even-adjoint}
\ee
where we introduce $\widehat{\mathcal {E}}^d_\nu$, to regulate the physics. In general, the conformal generators are unbounded, but we would like to focus near the identity to discuss the Lie algebra.
Thus, the reflection-even generators remain skew-symmetric. Notice that by expanding around the identity, we can show that $A^{\dagger_{\rm OS}}=-\tau(A)$ for operator $A$.

The odd directions require a different argument because a generic
$Y\in\mathfrak q$ does not preserve the hemisphere.  First, let us  consider an odd
direction for which the positive orbit
$\exp(tY)$ lies in the compression semigroup.  Since $\tau(Y)=-Y$, we obtain
\be
\exp(tY)^\sharp
=\tau(\exp(tY))^{-1}
=\exp(-tY)^{-1}
=\exp(tY).
\ee
Thus, the infinitesimal generator is symmetric. 
The L\"uscher-Mack theorem is needed at this point because a generic
reflection-odd transformation does not preserve the unit ball and
therefore does not act directly on the OS quotient.  In the case of \cite{neeb2014reflection,neeb2018reflection}, the theorem
states that the strongly continuous $*$-representation of the
compression semigroup analytically continues to a unitary
representation of the simply connected $c$-dual group with Lie algebra
$\mathfrak h\oplus i\mathfrak q$.  It thereby reconstructs the remaining
odd generators o and ensures that they obey
the correct Lie brackets.  Positivity of the radial Hamiltonian is a
separate consequence of the contraction semigroup discussed in the
next subsection.
The input of the L\"uscher-Mack theorem has a direct physical meaning.
Conformal transformations that move insertions further into the ball
act continuously on the reconstructed states. Accordingly, successive geometric
transformations act successively also on the Hilbert space, and reflection corresponds to taking its Hilbert space adjoint
\[
\widehat U(g_1g_2)
=
\widehat U(g_1)\widehat U(g_2),
\quad 
\widehat U(g^\sharp)
=
\widehat U(g)^{\dagger_{\rm OS}}.
\]
Mathematically, these properties define a strongly continuous
$*$-representation of the compression semigroup. For details, please refer to the original proof \cite{neeb2014reflection,neeb2018reflection}.
 
 Using the theorem, the odd generators after OS quotient obey
\be
\ip{Yu}{v}_{\OS}=\ip{u}{Yv}_{\OS},
\quad 
Y\in\mathfrak q,
\quad 
u,v\in\widehat{\mathcal {E}}^d_\nu.
\label{eq:odd-adjoint}
\ee

The above discussions can be understood more from physics point of view.
A symmetric odd generator becomes skew-symmetric after multiplication by
$i$.  The real Lie algebra represented by skew-symmetric operators is
therefore
\be
\mathfrak g^c
=
\mathfrak h\oplus i\mathfrak q
\subset\mathfrak g_{\C}:=\mathfrak g\otimes_{\R}\C.
\label{eq:c-dual-def}
\ee
$\mathfrak g_{\C}$ is the  complexification of the de Sitter algebra.
This real subspace is closed under commutators because
\be
[\mathfrak h,i\mathfrak q]\subset i\mathfrak q,
\quad 
[i\mathfrak q,i\mathfrak q]
=
-[\mathfrak q,\mathfrak q]
\subset\mathfrak h.
\ee
The minus sign in the second relation is the algebraic origin of the change
of real form.

We can now identify this real algebra explicitly.  Let
$a,b=0,1,\ldots,d$ denote the embedding  coordinates not reversed by $R_0$.  A
generator $L_{ab}$ rotates or boosts only within this fixed
$(d+1)$-dimensional subspace and therefore  it is  even.  A generator
$L_{a,d+1}$ mixes a fixed coordinate with the reflected direction and it is
odd:
\be
\tau(L_{ab})=L_{ab},
\quad 
\tau(L_{a,d+1})=-L_{a,d+1}.
\ee
Hence,
\be
\mathfrak h\simeq\mathfrak{so}(1,d),
\quad 
\mathfrak q=\Span\{L_{a,d+1}\}.
\ee
According to \eqref{eq:c-dual-def}, the skew-symmetric generators of the
reconstructed representation are therefore
\be
\widetilde L_{ab}=L_{ab},
\quad 
\widetilde L_{a,d+1}=iL_{a,d+1}.
\label{eq:c-dual-generators}
\ee
The change of real form can be seen directly in their commutators.  The
$\widetilde L_{ab}$ obey the original $\mathfrak{so}(1,d)$ brackets, and
their brackets with $\widetilde L_{c,d+1}$ are unchanged.  The only sign that
changes is the bracket of two generators involving the reflected direction:
\be
[\widetilde L_{a,d+1},\widetilde L_{b,d+1}]
=-[L_{a,d+1},L_{b,d+1}]
=+L_{ab}.
\label{eq:odd-odd-sign}
\ee
In the original algebra the same bracket is $-L_{ab}$ because
$X^{d+1}$ is spacelike.  Equation~\eqref{eq:odd-odd-sign} is instead the
orthogonal-algebra commutator obtained when the metric component in the
$X^{d+1}$ direction is negative.  The effective embedding metric is thus
\be
\eta^c=\operatorname{diag}(-1,+1,\ldots,+1,-1),
\ee
which has two timelike and $d$ spacelike directions.  Therefore
\be
\mathfrak g^c\simeq\mathfrak{so}(2,d).
\label{eq:so2d}
\ee

  The geometric reflection does not turn the de Sitter
hyperboloid itself into AdS since it just induces $X_{d+1}\mapsto -X_{d+1}$.  Rather, the OS adjoint selects
$\mathfrak{so}(2,d)$ as a different real form of the common complexified
algebra $\mathfrak{so}(d+2,\C)$.
This real form has a familiar double interpretation: it is the isometry
algebra of $\AdS_{d+1}$ and the global conformal algebra of a
$d$-dimensional CFT.  The OS reconstruction selects a unitary positive-energy
representation of this algebra \cite{neeb2014reflection,neeb2018reflection}, which will be explained in the next section.

It remains to translate this result into the generators familiar from CFT.
From \eqref{eq:conformal-basis}, $M_{ij}=L_{ij}$ and
$L_{0i}=(P_i-K_i)/2$ are even, whereas
$D=L_{0,d+1}$ and $L_{d+1,i}=(P_i+K_i)/2$ are odd.  The even-odd adjoint
decomposition is equivalently
\be
\tau(D)=-D,
\quad 
\tau(M_{ij})=M_{ij},
\quad 
\tau(P_i)=-K_i,
\quad 
\tau(K_i)=-P_i.
\label{eq:tau-conformal-basis}
\ee
The adjoint rules therefore say that $M_{ij}$ and $P_i-K_i$ are skew-symmetric, while
$D$ and $P_i+K_i$ are symmetric. Using $A^{\dagger_{\rm OS}}=-\tau(A)$ for operator $A$, we recover the standard radial adjoint
\be
D^{\dagger_{\rm OS}}=D,
\quad 
P_i^{\dagger_{\rm OS}}=K_i,
\quad 
M_{ij}^{\dagger_{\rm OS}}=-M_{ij}.
\label{eq:radial-adjoint}
\ee
This is the precise sense in which the Bousso-Maloney-Strominger reflection \cite{Bousso:2001mw}
converts the de Sitter one-particle pairing into a radial-quantization
pairing. Similar construction in the Klein-Gordon inner product was discussed in \cite{Jafferis:2013qia}.  In particular, $D$ is now a candidate self-adjoint Hamiltonian and
translations are adjoint to special conformal transformations, as required
for a unitary CFT module.

   A related construction was discussed in \cite{Jafferis:2013qia}. In the static patch, the Klein–Gordon norm of quasinormal modes diverges at the cosmological horizon. To obtain a finite pairing and study completeness, the authors introduced an $R$-norm involving a spatial reflection closely related to the one used here. They also observed that the resulting adjoint relations select the AdS real form. Their $R$-norm, however, need not be positive for arbitrary choices of positive and negative-frequency modes. Clarifying its precise relation to the present OS quotient would be worthwhile.

\subsection{Radial evolution and positive energy}
\label{sec:radial-evolution}

The previous subsection identifies the reconstructed real form as
$\mathfrak{so}(2,d)$ and determines its radial adjoint relations.
However, the relation $D^\dagger_{\rm OS}=D$ alone does not imply that
the spectrum of $D$ is bounded from below.  We now use inward radial
evolution to establish this positive-energy property.

Introduce the inward radial-time coordinate $\rho=-\log r$, where
$r=|x|$.  Evolution by $t\geq0$ acts on the ball as
\be
\delta_t:x\longmapsto e^{-t}x,
\quad 
\delta_{t_1}\delta_{t_2}=\delta_{t_1+t_2}.
\label{eq:radial-contraction}
\ee
Since $\delta_t(B_1)\subset B_1$, these transformations belong to the
compression semigroup and act on the OS quotient by
\be
T_t[u]:=\widehat U(\delta_t)[u]=[U(\delta_t)u].
\label{eq:radial-semigroup}
\ee
Moreover, $\delta_t=\exp(-tD)$ and $\tau(D)=-D$ imply
\be
\delta_t^\sharp
=
\tau(\delta_t)^{-1}
=
\delta_t.
\label{eq:radial-sharp}
\ee

The Osterwalder-Schrader semigroup lemma is the step that converts
reflection positivity into positive energy.  It states that if a
 Euclidean evolution $U_t$ with $t\geq0$, preserves the
positive region and is reversed by reflection,
\be
\theta U_t\theta=U_{-t},
\ee
then its induced action on the OS quotient is a strongly continuous
self-adjoint contraction semigroup
\be
T_t^\dagger=T_t,\quad 
\norm{T_t}\leq1,\quad 
T_{t+s}=T_tT_s.
\ee
Consequently,
\[
T_t=e^{-tD_{\rm OS}},
\quad 
D_{\rm OS}\geq0.
\]
Physically, reflection positivity ensures that Euclidean evolution
can damp states but cannot amplify them, and therefore reconstructs a
nonnegative Hamiltonian.

It remains to identify the bottom of the spectrum.  Since the kernel $R_\nu(x,y)$ is positive semi-definite, we can define $R_\nu(x,y)=\braket{x|y}$. After taking the OS quotient, we can define similar basis $\ket{x}_{\rm OS}$ with positive definite space. Especially, $R_\nu(0,0)=1$ leads to the  localized normalizable basis near the center
denoted by $\ket{0}_{\rm OS}=\ket{\Delta_-}_{\rm OS}$.  The smeared states are constructed as
\be
W[\psi]=\int_{B_1}d^dx \,\psi\,(x)\,\ket{x}_{\rm OS}.
\ee
The time evolution is 
\be
T_tW[\psi]= W[U(\delta_t)\psi].
\ee
Since the wavefunction with weight $\Delta_+$ transforms as
\be
(U(\delta_t)\psi)(x)=e^{t\Delta_+}\psi(e^t x),
\ee
we find
\begin{align}
    T_tW[\psi]&=\int_{B_1}d^dx \,e^{t\Delta_+}\psi\,(e^tx)\,\ket{x}_{\rm OS}\nn\\
    &=e^{-t\Delta_-}\int_{B_1}d^dy \,\psi(y)\ket{e^{-t}y}_{\rm OS}.
\end{align}
Thus, the ket transforms as
\be
\widehat{U}(\delta_t)\ket{y}_{\rm OS}=e^{-t\Delta_-}\ket{e^{-t}y}_{\rm OS},
\ee
which leads to
\be
\widehat{U}(\delta_t)\ket{\Delta_-}_{\rm OS}=e^{-t\Delta_-}\ket{\Delta_-}_{\rm OS}.
\ee
By expanding it around $t=0$, we finally obtain
\be
D_{\rm OS}\ket{\Delta_-}_{\rm OS}
=
\Delta_-\ket{\Delta_-}_{\rm OS}.
\ee
The origin is fixed by rotations and special conformal
transformations.  Covariance of the kernel state therefore gives
\be
K_i\ket{\Delta_-}_{\rm OS}=0,
\quad 
M_{ij}\ket{\Delta_-}_{\rm OS}=0.
\label{eq:reconstructed-primary}
\ee
Translations generate the kernel states at other points, and their
span is dense in the reconstructed Hilbert space.  Hence the
reconstructed representation is the scalar positive-energy
lowest-weight module of $\mathfrak{so}(2,d)$ with lowest weight
$\Delta_-$.  With the opposite orientation of the compact energy
generator, the same structure is described as a negative-energy
highest-weight module.

\section{Relation to the Klebanov-Witten window}
\label{sec:kw}

We now compare the derived mass range with the mass of scalar field  in AdS.  We introduce  parameters for the
dS and AdS:
\be
\nu_{\dS}^{\,2}
=
\frac{d^2}{4}-m_{\dS}^2,
\quad
\nu_{\AdS}^{\,2}
=
\frac{d^2}{4}+m_{\AdS}^2,
\quad
\Delta_\pm
=
\frac d2\pm\nu_{\AdS}.
\label{eq:ds-ads-parameters}
\ee
We identify
\be
\nu_{\AdS}=\nu_{\dS}=:\nu.
\label{eq:nu-identification}
\ee
 This does not identify the dS and
AdS masses, nor does it analytically continue one Lorentzian spacetime
into the other.
For the principal series, $\nu=i\mu$ with $\mu>0$.  The formal AdS
dimensions are then complex and the corresponding mass lies below the
Breitenlohner-Freedman bound.  This is consistent with the earlier OS
result that the standard principal-series reflected form has a zero
quotient.  We therefore focus below on the complementary branch, for
which $\nu$ and $\Delta_\pm$ are real.

Near the AdS boundary, a scalar field has the asymptotic form
\be
\Phi(z,x)
\sim
z^{\Delta_-}\alpha(x)
+
z^{\Delta_+}\beta(x).
\label{eq:ads-falloffs}
\ee
In standard quantization, $\alpha$ corresponds to the source and the VEV of the dual
operator $\beta$ has dimension $\Delta_+$.  In alternative quantization,
$\beta$ is regarded as the source and the VEV of the operator $\alpha$ has the lower dimension
$\Delta_-$.

In our paper, we assumed that the basis has a weight $\Delta_-$.  The requirement from the positivity of the kernel requires 
\be
\Delta_-\geq\frac{d-2}{2}
\quad \Longleftrightarrow\quad 
\nu\leq1.
\label{eq:rp-unitarity}
\ee
This is the same as the unitarity bound in the CFT side. Correspondingly,  this is equivalent to the Klebanov-Witten window \cite{Klebanov:1999tb}, which describes the alternative quantization
\be
-\frac{d^2}{4}
<
m_{\AdS}^2
<
-\frac{d^2}{4}+1,
\label{eq:kw-mass-window}
\ee
where the lower bound is Breitenlohner-Freedman bound.

On the other hand, we suppose that the basis $\ket{\Delta_+}$ has a weight $\Delta_+$, corresponding to the standard quantization. The positivity of the reflected kernel suggests\footnote{In this case, if we naively repeat  the previous steps to construct the OS form, we need an additional factor to cancel $1/\Gamma(-\nu)$, which flips the sign depending on $\nu$.  This is because we have to use the inverse of the shadow kernel.  } 
\be
\Delta_+\geq\frac{d-2}{2}.
\ee
Since $\Delta_+=\frac{d}{2}+\nu$ with $\nu\geq0$, this is automatically satisfied.  The original UIR of de Sitter space has a shadow transformation and the two series $\ket{\Delta_+}$ and $\ket{\Delta_-}$ are unitary equivalent. The difference of the  the mass range in our case is caused by the additional reflection. 

For $d>2$, the point $\nu=1$ lies inside the open de Sitter
complementary series.  At this point $\Delta_-=(d-2)/2$ saturates the scalar
 unitarity bound and the level-two trace descendant
$P^2|\Delta_-\rangle_{\rm OS}$ has zero reflected norm.  Quotienting
by the null submodule generated by this descendant gives the short
scalar representation of $\mathfrak{so}(2,d)$.  This shortening is a
property of the OS-reconstructed module, not of the original de
Sitter complementary-series UIR. For $d=2$, $\nu=1$ corresponds to $\Delta_-=0$. This suggests $P_\mu\ket{0}=0$. Thus, after taking OS quotient, only the trivial state $\ket{0}$ remains. This is different from the higher-dimensional case since $P_\mu\ket{0}\neq0$.

\section{Conclusions }
\label{sec:conclusions}

We have studied equatorial reflection positivity for scalar unitary
representations of the connected de Sitter group.  The positivity tested in
this paper is additional to de Sitter unitarity: it is defined by composing
the original invariant Hilbert product with an equatorial reflection in  $S^d$  and restricting the wavefunctions to one
hemisphere.  The principal and complementary  series behave differently under
this test.  In the standard principal-series realization, locality makes the
reflected form vanish identically  and its Osterwalder-Schrader quotient is the zero
Hilbert space.  For the complementary series, the nonlocal shadow product is
instead converted into a nonzero  kernel on the unit ball.

Within the open complementary series, the Neeb-\'{O}lafsson classification
shows that this kernel is positive semidefinite precisely when
$\nu\leq1$.  The level-two trace descendant gives a direct physical
diagnosis of the same bound as the unitarity bound in CFT.  

The OS quotient also explains the emergence of the AdS real form.
Equatorial reflection changes the adjoint and splits the de Sitter algebra
into reflection-even and reflection-odd parts.  The fixed subgroup and the
conformal compression semigroup act on the positive region.  After
multiplying the odd generators by $i$, the resulting $c$-dual algebra is
$\mathfrak{so}(2,d)$.  Inward radial dilations are represented by a
self-adjoint contraction semigroup, whose generator is nonnegative, which supports the positivity of the dilatation operator.  The
nontrivial quotient is consequently a scalar positive-energy lowest-weight
module with lowest energy $\Delta_-$.  This results from a change of real form
and Hilbert-space structure; it is not an analytic continuation of the de
Sitter geometry into AdS.

Identifying the de Sitter and AdS spectral parameters,
$0<\nu<1$ leads to the Klebanov-Witten alternative-quantization window.  The
agreement has a simple representation-theoretic origin: the reflected shadow
product reconstructs the lower-weight  module, and reflection
positivity imposes the scalar conformal unitarity bound on that weight. 

The present construction is limited to a free scalar one-particle sector and
produces a single module of the global conformal algebra, not a complete
holographic dual or local conformal field theory.  A potential future direction will be the analysis with dynamical gravity towards understanding the holographic principle in dS. A related interesting technical problem is the Virasoro extension in the dS$_3$ case. The local Virasoro generators usually do not preserve the equator and we need some trick to repeat the techniques in \cite{neeb2014reflection,neeb2018reflection} and our procedure.

\section*{Acknowledgement}
We thank Y. Moriwaki for bringing the relevant mathematical theorem to our attention. We also thank Y. Moriwaki, T. Noumi, and K. Shinmyo for useful discussions during the early stages of this work. Y.S. is supported by the RIKEN Special Postdoctoral Researcher Program.
 OpenAI Codex (GPT-5.6-sol) was
used for the confirmation of computations in the manuscript.
The author is fully responsible for all scientific claims and
the manuscript.
\clearpage
\begin{appendices}

\section{Klein-Gordon product and boundary representation spaces}
\label{app:kg}

This appendix explicitly explains how the standard one-particle inner products arise from the bulk Klein-Gordon form. This is used to fix the notation rather than to use in the reflection positivity.  We set the de Sitter radius to $\ell=1$ in this appendix.

In global coordinates,
\be
\dd s^2=-\dd T^2+\cosh^2T\,\dd\Omega_d^2,
\ee
where $T$ is global de Sitter time and $\dd\Omega_d^2$ is the unit-round
metric on $S^d$.  On a constant-$T$ Cauchy surface, we can define the Klein-Gordon inner product
\be
(\phi^{(1)},\phi^{(2)})_{\KG}
=
i\cosh^dT
\int_{S^d}\dd\Omega\,
\left((\phi^{(1)})^*\partial_T\phi^{(2)}
-(\partial_T(\phi^{(1)})^*)\phi^{(2)}
\right),
\label{eq:KG-product}
\ee
which is not in general positive definite.

\subsection{Principal series}

For $m_{\dS}^2>d^2/4$, let us consider two solutions
$\phi^{(r)}$, $r=1,2$, and take the future limit $T\to+\infty$.  Let
$a^{(r)}(\Omega)$ and $b^{(r)}(\Omega)$ denote their two boundary
coefficients
\be
\phi^{(r)}(T,\Omega)
\sim
 a^{(r)}(\Omega)e^{-(d/2+i\mu)T}
+b^{(r)}(\Omega)e^{-(d/2-i\mu)T}.
\ee
The mixed $a^*b$ and $b^*a$ terms cancel in the antisymmetrized current,
while the $a^*a$ and $b^*b$ terms remain with opposite signs.  Substitution
into \eqref{eq:KG-product} therefore gives, up to a positive overall factor
that depends on conventions,
\be
(\phi^{(1)},\phi^{(2)})_{\KG}
\propto
\mu\int_{S^d}\dd\Omega\,
\left((a^{(1)})^*a^{(2)}-(b^{(1)})^*b^{(2)}\right).
\label{eq:KG-principal}
\ee
The full solution space has an indefinite Klein-Gordon form.  A one-particle
Hilbert space is obtained by choosing a positive-frequency subspace.  At the
level of asymptotic representation data, the simple choice $b=0$ gives an
$L^2(S^d)$ norm.  More generally, one may impose $b=\mathcal R a$, where
$\mathcal R$ is a bounded operator on $L^2(S^d)$ satisfying the strict
operator inequality $1-\mathcal R^\dagger\mathcal R>0$.  Here
$\mathcal R^\dagger$ denotes the adjoint with respect to the standard
$L^2(S^d)$ inner product.  The inequality selects a positive subspace.  The choice of positive-frequency subspace is vacuum data.

\subsection{Complementary series}

For a parameter $0<\nu<d/2$ with
$\nu\notin\mathbb N$, where $\mathbb N=\{1,2,\ldots\}$, let us consider two
solutions $\phi^{(r)}$, $r=1,2$.  Their two leading future falloffs are
\be
\phi^{(r)}(T,\Omega)
\sim
f_+^{(r)}(\Omega)e^{-(d/2+\nu)T}
+
f_-^{(r)}(\Omega)e^{-(d/2-\nu)T}.
\ee
  Because the
falloff exponents are real, the two same-falloff contributions cancel in the
antisymmetrized current.   Up to a positive convention-dependent
constant, the Klein-Gordon form becomes
\be
(\phi^{(1)},\phi^{(2)})_{\KG}
\propto
i\nu
\int_{S^d}\dd\Omega\,
\left((f_+^{(1)})^*f_-^{(2)}-(f_-^{(1)})^*f_+^{(2)}
\right).
\label{eq:KG-complementary}
\ee
Unlike the principal series, the two asymptotic coefficients carry real,
shadow-related weights.  The fast coefficient $f_+$ has weight $\Delta_+$,
while the slow coefficient $f_-$ has weight $\Delta_-$.  The subscripts now
agree with the corresponding falloff exponents and avoid the conventional
AdS symbols $\alpha$ and $\beta$ used in Section~\ref{sec:kw}.  The kernel
with exponent $\Delta_-$ is therefore the shadow intertwiner from the fast
coefficient to the slow one.  A de Sitter-invariant positive-frequency
subspace relates them as follows.  Suppressing the solution label $r$,
\be
f_-=c_\nu S_\nu f_+,
\quad 
(S_\nu f_+)(\Omega)
=
\int_{S^d}\dd\Omega'\,
\frac{f_+(\Omega')}{(1-\Omega\cdot\Omega')^{\Delta_-}},
\label{eq:shadow-map}
\ee
where $c_\nu$ is chosen so that the resulting form is positive.  The
shadow operator $S_\nu$ is self-adjoint and positive with respect to the
round-sphere $L^2$ pairing in the complementary range.
Substituting this relation into the asymptotic Klein–Gordon form gives
\be
(\phi^{(1)},\phi^{(2)})_{\KG}
\propto
-\operatorname{Im}(c_\nu)
\int_{S^d}\dd\Omega
\int_{S^d}\dd\Omega'\,
(f_+^{(1)}(\Omega))^*
(1-\Omega\cdot\Omega')^{-\Delta_-}
f_+^{(2)}(\Omega'),
\ee
which is the complementary-series shadow inner product \eqref{eq:comp-product} on wavefunctions of weight $\Delta_+$.  This is the original positive de Sitter inner product.  The OS form studied in the main text is obtained only after inserting the additional reflection $r_0$. 

\end{appendices}

\bibliographystyle{utphys}
\bibliography{dSreflection}
\end{document}